\documentclass{appolb}
\usepackage{graphicx}
\usepackage{amssymb}
\usepackage{amsmath}
\usepackage{hyperref}
\usepackage{cleveref}
\usepackage{cite}
\headtitle{Precise calculations of Higgs boson decays Printed on \today}
\headauthor{Precise calculations of Higgs boson decays Printed on \today}

\begin{document}
\title{Precise calculations for decays of Higgs bosons in extended Higgs sectors%
\thanks{Presented by Kodai Sakurai at XXX Cracow Epiphany Conference on Precision Physics at High Energy Colliders,
Cracow, Poland, January 8--12, 2024}%
}

\author{
Kodai Sakurai
\footnote{\href{kodai.sakurai.e3@tohoku.ac.jp}{kodai.sakurai.e3@tohoku.ac.jp},}
\address{Institute of Theoretical Physics, Faculty of Physics, University of Warsaw, ul. Pasteura 5, PL-02-093 Warsaw, Poland}
\address{Department of Physics, Tohoku University, Sendai, Miyagi 980-8578, Japan}}

\maketitle
\begin{abstract}
We briefly introduce \verb|H-COUP_3.0|, which we developed for evaluating higher-order corrections to any Higgs boson decays in various extended Higgs sectors.
Focusing on two Higgs doublet models (2HDMs), we then discuss how the non-decoupling effects of the additional Higgs bosons are significant in Higgs boson decays.
\end{abstract}
  
\section{Introduction}
The Higgs boson with a mass of 125 GeV has been discovered in the LHC experiments and afterward, more data for the measurements of the Higgs boson have been accumulated~\cite{ATLAS:2022vkf, CMS:2022dwd}. 
The recent results indicate that, within the current experimental and theoretical uncertainties, the discovered Higgs boson has similar properties to the one predicted in the Standard Model (SM). 
However, this does not mean the possibility of the extended Higgs sector is ruled out since one can take a so-called alignment limit for Higgs boson states or consider a scenario close to this limit. 
In such a limit, the predictions of the Higgs couplings are close to those of the SM, so that extended Higgs models can fit the current measurements of the Higgs boson.  
The importance of the extended Higgs sector is that it often appears in a variety of new physics models that explain phenomena beyond the SM such as tiny neutrino masses, the existence of dark matter, and the baryon asymmetry of the universe. 
Hence, by investigating the real shape of the Higgs sector, one can approach new physics beyond the SM. 

Determination of the shape of the Higgs sector can be achieved in two independent ways, i.e., the direct searches of additional Higgs bosons and precision measurements of the 125 GeV Higgs boson. The former gives direct evidence of the existence of the additional Higgs bosons and has been performed at the LHC experiments. 
Although the current reach for the mass range of the additional Higgs bosons is below TeV, a higher energy range can be survived by, 
e.g., High-Luminosity LHC (HL-LHC)~\cite{Azzi:2019yne}, a 100 TeV proton-proton collider (FCC-hh)~\cite{FCC:2018byv}, and muon colliders~\cite{Accettura:2023ked}. 
For the latter, due to a mixing among Higgs bosons and higher order corrections, the predictions for Higgs observables, e.g., Higgs couplings, the decay rates, and the production cross sections, typically deviate from the SM. 
The pattern of the deviations, meaning how the deviations in different Higgs observables correlate, can be characteristic depending on extended Higgs models. 
Such a deviation can be detected in future precision measurements such as HL-LHC~\cite{Azzi:2019yne}, the International Linear Collider (ILC)~\cite{Fujii:2019zll}, the Circular Electron-Positron Collider (CEPC)~\cite{CEPC-SPPCStudyGroup:2015csa} and $e^+e^-$ collisions of the Future Circular Collider (FCC-ee)~\cite{FCC:2018byv}. 
By combining these different approaches, one can explore a large portion of parameter space in extended Higgs sectors~\cite{Aiko:2020ksl}. 

The future measurements of the 125 GeV Higgs boson are expected to be performed at the accuracy of \% level or less. 
At this level of precision, to compare such precise data with theoretical predictions, it is essential to evaluate not only QCD corrections but also higher-order corrections arising from electroweak and Higgs interactions in theoretical calculations. 
Furthermore, the inclusion of higher-order corrections is also important for direct searches of the additional Higgs bosons. 
The reason is that the current data of the 125 GeV Higgs boson favors exact alignment and/or near alignment scenarios.  
In these scenarios, the Higgs-to-Higgs decays, e.g., $H\to hh, A\to Zh$ in 2HDMs, are suppressed by the Higgs mixing parameters at the tree level. 
Thus, the higher-order corrections to these decay modes become significant. 

We have studied the next leading order (NLO) EW corrections to the decays of 125 GeV Higgs bosons~\cite{Kanemura:2004mg, Kanemura:2014dja, Kanemura:2015mxa, Kanemura:2015fra, Kanemura:2016sos, Kanemura:2016lkz, Kanemura:2017wtm, Kanemura:2018yai, Kanemura:2019kjg} and additional Higgs bosons~\cite{Kanemura:2022ldq, Aiko:2022gmz, Aiko:2021can} in the context of the extended Higgs models in the previous works.
Relating them, we have developed {\tt H-COUP}~\cite{Kanemura:2017gbi, Kanemura:2019slf, Aiko:2023xui}, a public tool to compute the higher-order corrections to the 125 GeV Higgs and additional Higgs boson decays in various extended Higgs models. 
In the current version (\verb|H-COUP_3.0|), the 2HDMs with softly broken $Z_2$ symmetry, the Higgs singlet model, and the inert doublet model are implemented. 
Features of the H-COUP are the following:
1.~One can calculate the theoretical predictions for the decays of any Higgs bosons and easily compare them among the different extended Higgs models.  
2.~One-loop corrections to vertex functions for the 125 GeV Higgs boson can be computed in any momentum of external particles.  
Apart from {\tt H-COUP}, similar public tools are also available, e.g., {\tt 2HDECAY}~\cite{Krause:2018wmo}, {\tt Prophecy4f}~\cite{Denner:2019fcr}, {\tt ewN2HDECAY}~\cite{Krause:2019oar}, {\tt EWsHDECAY}~\cite{Egle:2023pbm}, and~{\tt FlexibleDecay}~\cite{Athron:2021kve}. 
In the following, we give an overview of \verb|H-COUP| in Sec.~2. 
We then mainly focus on the 2HDMs as a concrete example of the simple extended Higgs models and give the physical applications of the {\tt H-COUP} program in Sec.~3. The summary is given in Sec.~4.  



\section{Overview of {\tt H-COUP}}

\verb|H-COUP_3.0| can be downloaded from \url{http://www-het.phys.sci.osaka-u.ac.jp/~hcoup}.
This is the numerical computation program to evaluate the NLO EW corrections to the 125 GeV Higgs boson decays and additional Higgs bosons in the extended Higgs models. 
QCD corrections are also computed up to NLO or NNLO depending on the decay processes. 
In the current version, the following models are implemented, 
Four types of 2HDMs (Type I, Type II, Type X, and Type Y)~\cite{Kanemura:2004mg}, 
Higgs singlet model (HSM)~\cite{Kanemura:2015fra}, 
Inert doublet model (IDM)~\cite{Kanemura:2016sos}. 

Higher-order corrections are calculated in the following processes: \par\noindent
\noindent for the 125 GeV Higgs boson$(h)$, 
\begin{equation} 
    \label{eq:hdec}
    h\to f\bar{f},~h\to ZZ^\ast\to Zf\bar{f},~ h\to WW^\ast\to Wf\bar{f^\prime},~
    h\to gg/\gamma\gamma/Z\gamma\;,    
\end{equation}
for the CP-even Higgs boson $H$, CP-odd Higgs boson $A$, and charged Higgs bosons $H^\pm$, 
\begin{align}\label{eq:bhdec}
    &H\to f\bar{f}\;,~~H\to ZZ\;,~~ H\to WW\;,~~H\to AZ\;,~~ H\to W^\pm H^\mp\;,  \notag\\ 
    &H\to hh\;,~~ H\to AA \;,~~ H\to H^\pm H^\mp\;,~~ H\to gg/\gamma\gamma/Z\gamma\;, \notag\\ 
    &A\to f\bar{f},~A\to hZ,~ A\to HZ,~ A\to W^\pm H^\mp ,~
    A\to gg/\gamma\gamma/Z\gamma,\notag\\ 
    &H^\pm\to f\bar{f^\prime},~~H^\pm\to W^\pm h \;,~~H^\pm\to W^\pm H \;,~~ H^\pm\to W^\pm A \;. 
\end{align}
Except for the loop-induced decays ($h/H/A\to gg/\gamma\gamma/Z\gamma$), full NLO EW corrections to all the above processes are evaluated, and QCD corrections
are also calculated where applicable. 
The loop-induced decays are evaluated at EW LO, but the QCD corrections are calculated~\footnote{Other loop-induced decays such as $A\to ZZ/WW$ and $H^\pm\to W^\pm\gamma/W^\pm Z$ are evaluated at LO.}. 
All the decay modes, Eqs.~\eqref{eq:hdec} and \eqref{eq:bhdec}, happen in the 2HDMs while $A$ and $H^\pm$ do not exist in the HSM. 
For the IDM, scalar to scalar gauge type decay modes ($S\to SV$) only exist. 
Other decay processes are forbidden due to unbroken discrete $Z_2$ symmetry. 
\verb|H-COUP_3.0| outputs not only the decay rates of the Higgs bosons but also one-loop corrected vertex functions for $h$, 
\begin{equation}
\hat{\Gamma}_{hff}^i,~\hat{\Gamma}_{hZZ}^j,~\hat{\Gamma}_{hWW}^j,~\hat{\Gamma}_{hhh}~(i=S,P,V_1,V_2,A_1,A_1,T,PT,~j=1,2)
\end{equation}
in any momentum of external particles. 
The convention of the form factor decomposition can be found in Ref.~\cite{Kanemura:2017gbi}.

The renormalization of the Higgs sector, which is required in the evaluation of the NLO EW corrections, is mainly performed by the on-shell scheme. 
While the detailed discussion can be seen in Refs.~\cite{Kanemura:2017wtm, Aiko:2023xui}, we here highlight the features of the renormalization scheme used in the program below. 
For the renormalization of Higgs mixing parameters, it is pointed out that the on-shell renormalized mixing angles are gauge-dependent~\cite{Yamada:2001px}. 
To remove this unwanted gauge dependence, the pinch technique~\cite{Papavassiliou:1994pr} is applied. 
The counter terms of the mixing angles involve appropriate pinch term contributions and thereby they are gauge-independent. 
For the renormalization of the tadpoles, two different renormalization schemes, standard tadpole scheme (STS)~\cite{Hollik:1988ii} and alternative tadpole scheme (ATS)~\cite{Fleischer:1980ub}, are implemented. 
Some of the model input parameters are renormalized in the $\overline{\rm MS}$ scheme, 
e.g., in the 2HDMs the mass parameter $M^2$ corresponds to it. 
Since $M^2$ is defined by the softly broken $Z_2$ parameter $m_{12}$ though $M^2=m_{12}^2/{\cos\beta \sin\beta}$, 
one can also renormalize $m_{12}$ instead of $M$. 
Hence for the 2HDMs, combining the tadpole renormalization, four scheme options are available: 
1. STS, $\delta M^2$\;,~2. STS, $\delta m^2_{12}$\;,~3. ATS, $\delta M^2$\;,~and 4. ATS, $\delta m^2_{12}$.
This scheme difference is only relevant for the scalar decays into two scalar bosons ($S\to SS$) and the $hhh$ vertex function. It is shown that 
scheme 3 and scheme 4 give the same results for these quantities in Appendix~A of Ref.~\cite{Aiko:2023xui}. 
For the HSM, this type of redefinition of the model parameter is not carried out, so that the two scheme options for the tadpoles are available. 
For the IDM, there is no scheme difference between STS and ATS for the above-mentioned quantities and thus, the Higgs boson decays, and the renormalized vertex functions are computed in STS. 

\section{Radiative corrections to the Higgs boson decays}

In this section, we focus on the 2HDMs with softly broken $Z_2$ symmetry as a concrete example of simple extended Higgs models and illustrate the impact of radiative corrections to the Higgs boson decays. 
A detailed description of the model and our convention can be seen in, e.g., Ref.~\cite{Kanemura:2004mg}. 
The model involves two Higgs doublet fields with hypercharge $Y=1/2$ and each component field mix.  
CP-odd and charged Higgs fields are commonly diagonalized by a mixing angle $\beta$. 
The mixing angle for CP-even Higgs fields is parameterized by $\alpha$. 
In the mass basis, there are four additional Higgs bosons, the CP-even Higgs boson ($H$), the CP-odd Higgs boson ($A$), the charged Higgs bosons ($H^\pm$), in addition to the 125 GeV Higgs boson ($h$).
While \verb|H-COUP_3.0| can evaluate all the decays of Higgs bosons listed in Eqs.~\eqref{eq:hdec} and \eqref{eq:bhdec}, we here mainly discuss NLO EW corrections to the decays of $h$ and $A$. 

As already mentioned in the introduction, the measurements of $h$ and direct searches of the additional Higgs bosons restrict the model parameter spaces, especially for the mixing parameters $\cos(\beta-\alpha)$, $\tan\beta$ and the additional Higgs boson masses, $m_H$, $m_A$ and $m_{H^\pm}$. 
On the other hand, flavor experiments give independent constraints. 
In particular, $B \to X_s \gamma$ gives lower bounds for the charged Higgs boson masses, e.g., $m_{H^\pm}\gtrsim 600{\rm GeV}$ for Type II and Type Y~\cite{Misiak:2017bgg}. 
In the following, taking into account all the experimental constraints, we consider two distinct scenarios for the masses of the additional Higgs bosons:  
\begin{equation}
\mbox{Scenario~A: }m_A=m_H=300{\rm GeV},~\mbox{Scenario~B: }m_A=m_H=800{\rm GeV}.    
\end{equation}

\begin{figure}
 \centering
 \includegraphics[width=60mm]{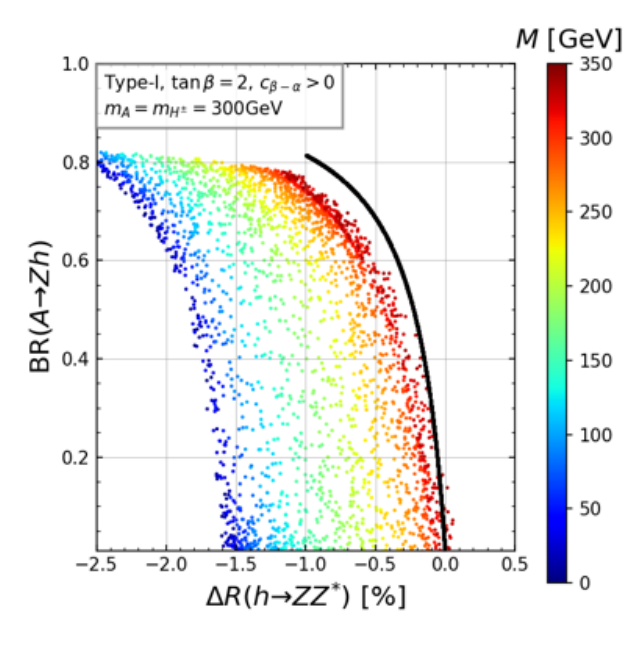}\vspace{0cm}
  \includegraphics[width=60mm]{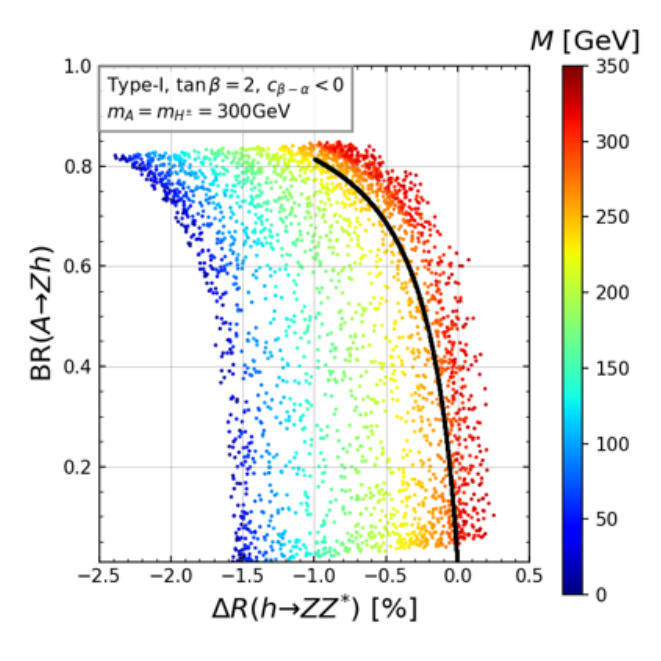}\vspace{0cm}
 \caption{The correlation between ${\rm BR}(A\to Zh)$ and the deviation in $h\to ZZ^\ast$ from the SM prediction, which is defined in Eq.~\eqref{eq:DR}, is shown in the Type~I 2HDM~\cite{Aiko:2022gmz}.}  
  \label{fig:corr}
\end{figure}

Scenario~A corresponds to the case where one has a relatively light mass spectrum of the additional Higgs bosons. 
Except for Type I, this scenario is excluded by the above-mentioned limit from $B \to X_s\gamma$~\cite{Misiak:2017bgg}. 
Hence, here we focus on the Type I 2HDM and discuss the effect of the higher order corrections to $A\to Zh$.  
Although in the alignment limit $\cos(\beta-\alpha)=0$ the tree level contributions to this decay mode are zero, 
the 1-loop corrections are not necessarily suppressed by $\cos(\beta-\alpha)$. 
Furthermore, at the loop level, the predicted Higgs boson couplings do not necessarily coincide with the SM predictions in the alignment limit because the nondecoupling effects of the additional Higgs bosons give a deviation from the SM predictions. 
Thus, the correlation between $A\to Zh$ and the deviation of the Higgs boson couplings can be changed much from the tree level analysis. 
This is illustrated in Fig.~\ref{fig:corr}. 
The horizontal axis is chosen by the deviation of the decay rate of $\Gamma(h\to ZZ^\ast)$, which is quantified by 
\begin{align}\label{eq:DR}
    \Delta R(h\to XX)= \frac{\Gamma(h\to XX)^{\rm 2HDM}}{\Gamma(h\to XX)^{\rm SM}}- 1
\end{align}
with $\Gamma(h\to XX)^{\rm 2HDM}$ ($\Gamma(h\to XX)^{\rm SM}$) being the decay rate of $h$ in the 2HDM (SM). 
While the color points show the results at NLO, the black lines show the tree level result. 
The remarkable behaviors that can be read from the figure are the following two things. 
For one thing, there is parameter space where $\Delta R(h\to ZZ^\ast)$ deviates with ${\cal O}(1)$\% due to the non-decoupling effects of the additional Higgs bosons but ${\rm BR}(A\to Zh)$ is close to zero.
For another thing, in the case of $\cos(\beta-\alpha)<0$, NLO corrections can increase both of $\Delta R(h\to ZZ^\ast)$ and ${\rm BR}({A\to Zh})$, so that $\Delta R(h\to ZZ^\ast)$ can close to zero while having ${\cal O}(10)\%$ of ${\rm BR}(A\to Zh)$. 
As seen from the figure, these behaviors are not realized by the tree level analysis. 
This result clearly shows a correlation between the decay properties of the additional Higgs boson and the observables of $h$ can be different from those indicated from the tree level results as long as the additional Higgs bosons have a non-decoupling feature. 

\begin{figure}
 \centering
  \includegraphics[width=60mm]{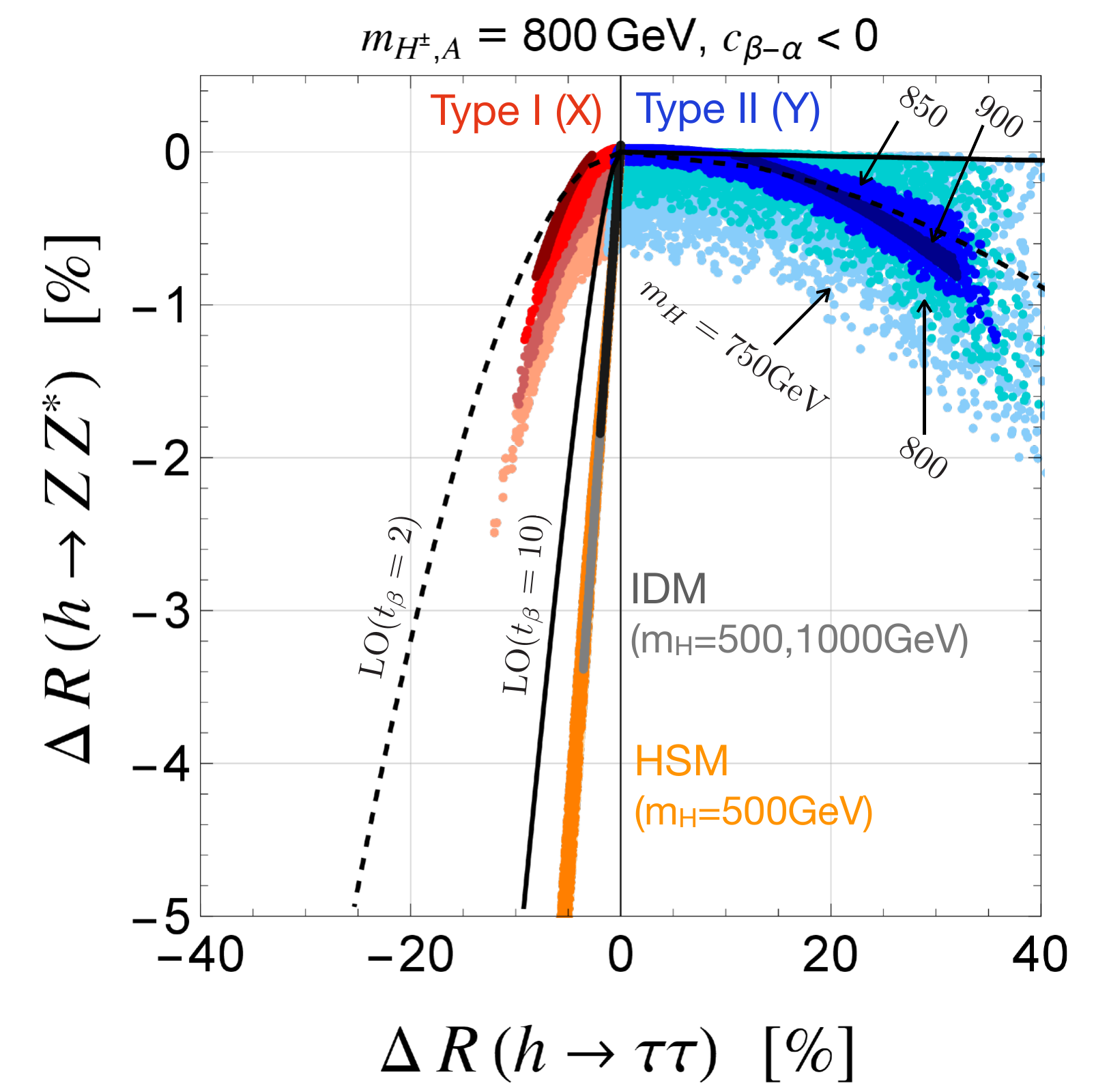}\vspace{0cm}
   \caption{ Pattern of the deviations from the SM prediction in the decays of the 125 GeV Higgs boson is shown in the plane of $\Delta R(h\to \tau\tau)$ and $\Delta R(h\to ZZ^\ast)$ for various extended Higgs models. The definition of $\Delta R$ is in Eq.~\eqref{eq:DR}.
 }  
  \label{fig:3}
\end{figure}

Scenario B corresponds to the case where one has a heavier mass spectrum of the additional Higgs bosons. 
All types of the 2HDMs can satisfy the $B\to X_s\gamma$ bounds, so that one can compare results among four types of the 2HDMs. 
The theoretical behavior of $A\to Zh$ in this scenario is presented in Ref.~\cite{Aiko:2022gmz}. 
Here we discuss how one of the four types of the 2HDMs can be separated from the other extended Higgs models by looking at the patterns of deviations in the decay rates of $h$. 
To demonstrate this we perform the scan analysis, imposing the theoretical constraints such as the unitarity at the tree level, vacuum stability, and wrong vacuum conditions. 
Furthermore, the bound from the electroweak oblique parameters is imposed as an experimental constraint. 
All of them are implemented in \verb|H-COUP_3.0|. 
For the 2HDMs, we take the input parameters as:
\begin{align}
    &m_A=m_{H^\pm}=800~{\rm GeV},\;m_H=750,\;800,\;850,\;900~{\rm GeV},~\cos(\beta-\alpha)<0, \notag\\
    &0.995<\sin(\beta-\alpha)<1\;,~ 2<\tan\beta<10\;,~ 0<M<m_H+500{\rm GeV}. 
\end{align}
We also analyze the HSM and IDM. We choose input parameters for them in the following range,
\begin{align}
\mbox{HSM:~}~&m_H=500{\rm GeV}\;,~\lambda_s=0.1,\;,~\mu_s=0\;,~ \notag \\
            &0.95<\cos\alpha<1\;, 0<m_s<m_H+500{\rm GeV},\\
\mbox{IDM:~}~&m_H=m_A=m_{H^\pm}=500,1000{\rm GeV}\;,\notag \\
& 0<\lambda_2<4\pi,~ 0<\mu_2<m_H+500{\rm GeV},
\end{align}
where definitions of these parameters are given in Ref.~\cite{Aiko:2023xui}.

The correlations between $\Delta R(h\to ZZ^\ast)$ and $\Delta R(h\to \tau\tau)$ for the 2HDMs, HSM and IDM are shown in Fig.~\ref{fig:3}. 
While the color points are the result including NLO EW corrections and NNLO QCD corrections, black solid and dashed lines show the results at the tree level with $\tan\beta=2$ and 10, in the 2HDMs respectively. One can see that the direction of the deviations is different depending on the models. 
For instance, $h\to \tau\tau$ deviates in the negative direction for the Type I and Type X 2HDMs. 
It is the opposite in the Type II and Type Y 2HDMs. 
In addition, one holds $\Delta R(h\to \tau\tau)\sim \Delta R(h\to ZZ^\ast)$ for the HSM and the IDM but this is not the case for the Type I and Type X. 
Type I and Type X can be distinguished by the correlation between $h\to \tau\tau$ and $h\to bb$~\cite{Kanemura:2018yai}.  
Although such a global picture of the pattern of the deviations in the $h$ decays is obtained even at the tree level, the NLO corrections generate qualitative differences. 
In the type II and the type Y, $\Delta R(h\to ZZ)$ can reach $-2\%$ in $\Delta R_{\tau\tau}\lesssim 40\%$. 
However, this shrinks to $-1\%$ in the tree level analysis. 
Hence, NLO corrections can increase $\Delta R (h\to ZZ^\ast)$ by 1\% even in the regime of heavier masses of the additional Higgs bosons. 
We note the correlation between $\Delta R({h\to WW^\ast})$ and $\Delta R({h\to \tau\tau})$ is similar to Fig.~\ref{fig:3}. 
The future sensitivity of $h\to WW^\ast$ can be $1.6\%$ at ILC with $2~{\rm ab}^{-1}$ of data, which is estimated from the measurement accuracy with $250~{\rm fb}^{-1}$ of data at 250 GeV in Ref.~\cite{Barklow:2017suo} by rescaling the luminosity. 
Therefore, the NLO corrections to the decay rates of $h$ are comparable with the future precise measurements. 

\section{Summary}
The extended Higgs sectors can be tested by precision measurements of the 125 GeV Higgs boson ($h$) and/or the discovery of additional Higgs bosons.
To compare with future precise measurements of $h$, higher-order corrections to observables of $h$ should be evaluated. 
It would also be important for direct searches of the additional Higgs bosons since the current measurements of $h$ favor a near-alignment scenario. 
We gave an overview of \verb|H-COUP|, the program to evaluate higher order corrections to decays of any Higgs bosons in various extended Higgs models and 
showed physical applications of this program. 
We have seen that the correlation between ${\rm BR}(A\to Zh)$ and $\Delta R(h\to ZZ^\ast)$ (defined by Eq.~\eqref{eq:DR}) can be different from the tree level results. 
Furthermore, due to the NLO EW corrections, {\cal O}(1)\% increase can be seen in $|\Delta R(h\to ZZ^\ast)|$ even if the additional Higgs bosons are relatively heavy.

\smallskip

\textbf{Acknowledgements} This work is supported by JSPS KAKENHI Grant No.~23KJ0086 and the National Science Centre, Poland, under research Grant No. 2020/38/E/ST2/00243. 
\bibliographystyle{utphys}
\bibliography{references} 

\end{document}